\documentclass[floats,floatfix,showpacs,amsmath,amssymb,aps,twocolumn,superscriptaddress,nofootinbib,nolongbibliography,prd]{revtex4-1}

\usepackage{amssymb,amsmath,verbatim,mathtools,needspace,enumitem,etoolbox,graphicx,physics,microtype,afterpage,xspace,tabularx,lmodern,multirow}

\usepackage{gensymb}
\usepackage{subcaption}
\usepackage{ragged2e}
\usepackage[dvipsnames, usenames]{xcolor}
\definecolor{linkcolor}{rgb}{0.0,0.3,0.5}
\usepackage[unicode, colorlinks=true, linkcolor=linkcolor, citecolor=linkcolor, filecolor=linkcolor, urlcolor=linkcolor, linktocpage, breaklinks]{hyperref}
\usepackage[all]{hypcap}
\usepackage{cancel}
\usepackage[T1]{fontenc}
\usepackage[utf8]{inputenc}
\usepackage[usenames,dvipsnames]{xcolor}
\usepackage{csquotes}
\hypersetup{colorlinks=true,citecolor=Periwinkle,linkcolor=Periwinkle,urlcolor=Periwinkle}

\definecolor{romared}{RGB}{142,0,28}

\newcommand{\be}{\begin{equation}}
\newcommand{\ee}{\end{equation}}

\def\be{\begin{equation}}
\def\ee{\end{equation}}
\newcommand{\beq}{\begin{eqnarray}}
\newcommand{\eeq}{\end{eqnarray}}
\usepackage{makecell}
\usepackage{soul}

\newcolumntype{Y}{>{\centering\arraybackslash}X}

\makeatletter
\newcommand*{\rom}[1]{\expandafter\@slowromancap\romannumeral #1@}
\makeatother

\makeatletter
\let\jnl@style=\rm
\def\ref@jnl#1{{\jnl@style#1}}

\def\aj{\ref@jnl{AJ}}                   
\def\actaa{\ref@jnl{Acta Astron.}}      
\def\araa{\ref@jnl{ARA\&A}}             
\def\apj{\ref@jnl{ApJ}}                 
\def\apjl{\ref@jnl{ApJ}}                
\def\apjs{\ref@jnl{ApJS}}               
\def\ao{\ref@jnl{Appl.~Opt.}}           
\def\apss{\ref@jnl{Ap\&SS}}             
\def\aap{\ref@jnl{A\&A}}                
\def\aapr{\ref@jnl{A\&A~Rev.}}          
\def\aaps{\ref@jnl{A\&AS}}              
\def\azh{\ref@jnl{AZh}}                 
\def\baas{\ref@jnl{BAAS}}               
\def\bac{\ref@jnl{Bull. astr. Inst. Czechosl.}}
\def\caa{\ref@jnl{Chinese Astron. Astrophys.}}
\def\cjaa{\ref@jnl{Chinese J. Astron. Astrophys.}}
\def\icarus{\ref@jnl{Icarus}}           
\def\jcap{\ref@jnl{J. Cosmology Astropart. Phys.}}
\def\jrasc{\ref@jnl{JRASC}}             
\def\memras{\ref@jnl{MmRAS}}            
\def\mnras{\ref@jnl{MNRAS}}             
\def\na{\ref@jnl{New A}}                
\def\nar{\ref@jnl{New A Rev.}}          
\def\pra{\ref@jnl{Phys.~Rev.~A}}        
\def\prb{\ref@jnl{Phys.~Rev.~B}}        
\def\prc{\ref@jnl{Phys.~Rev.~C}}        
\def\prd{\ref@jnl{Phys.~Rev.~D}}        
\def\pre{\ref@jnl{Phys.~Rev.~E}}        
\def\prl{\ref@jnl{Phys.~Rev.~Lett.}}    
\def\pasa{\ref@jnl{PASA}}               
\def\pasp{\ref@jnl{PASP}}               
\def\pasj{\ref@jnl{PASJ}}               
\def\rmxaa{\ref@jnl{Rev. Mexicana Astron. Astrofis.}}%
\def\qjras{\ref@jnl{QJRAS}}             
\def\skytel{\ref@jnl{S\&T}}             
\def\solphys{\ref@jnl{Sol.~Phys.}}      
\def\sovast{\ref@jnl{Soviet~Ast.}}      
\def\ssr{\ref@jnl{Space~Sci.~Rev.}}     
\def\zap{\ref@jnl{ZAp}}                 
\def\nat{\ref@jnl{Nature}}              
\def\iaucirc{\ref@jnl{IAU~Circ.}}       
\def\aplett{\ref@jnl{Astrophys.~Lett.}} 
\def\apspr{\ref@jnl{Astrophys.~Space~Phys.~Res.}}
\def\bain{\ref@jnl{Bull.~Astron.~Inst.~Netherlands}} 
\def\fcp{\ref@jnl{Fund.~Cosmic~Phys.}}  
\def\gca{\ref@jnl{Geochim.~Cosmochim.~Acta}}   
\def\grl{\ref@jnl{Geophys.~Res.~Lett.}} 
\def\jcp{\ref@jnl{J.~Chem.~Phys.}}      
\def\jgr{\ref@jnl{J.~Geophys.~Res.}}    
\def\jqsrt{\ref@jnl{J.~Quant.~Spec.~Radiat.~Transf.}}
\def\memsai{\ref@jnl{Mem.~Soc.~Astron.~Italiana}}
\def\nphysa{\ref@jnl{Nucl.~Phys.~A}}   
\def\physrep{\ref@jnl{Phys.~Rep.}}   
\def\physscr{\ref@jnl{Phys.~Scr}}   
\def\planss{\ref@jnl{Planet.~Space~Sci.}}   
\def\procspie{\ref@jnl{Proc.~SPIE}}   

\makeatother

\begin{document}

\title{Impact of Correlations on the Modeling and Inference of Beyond Vacuum-GR Effects in Extreme-Mass-Ratio Inspirals}

\author{Shubham Kejriwal} 
\email[]{shubhamkejriwal@u.nus.edu}
\affiliation{Department of Physics, National University of Singapore, Singapore 117551}
\author{Lorenzo Speri} 
\email[]{lorenzo.speri@aei.mpg.de}
\affiliation{Max Planck Institute for Gravitational Physics (Albert Einstein Institute), D-14476 Potsdam, Germany}
\author{Alvin J. K. Chua}
\email[]{alvincjk@nus.edu.sg}
\affiliation{Department of Physics, National University of Singapore, Singapore 117551}
\affiliation{Department of Mathematics, National University of Singapore, Singapore 119076}

\begin{abstract}
In gravitational-wave astronomy, extreme-mass-ratio-inspiral (EMRI) sources for the upcoming LISA observatory have the potential to serve as high-precision probes of astrophysical environments in galactic nuclei, and of potential deviations from general relativity (GR). Such ``beyond vacuum-GR'' effects are often modeled as perturbations to the evolution of vacuum EMRIs under GR. Many studies have reported unprecedented constraints on these effects by examining the inference of one effect at a time. However, in a more realistic analysis, the simultaneous inference of multiple such effects is required since the parameters describing them are generally significantly correlated with each other and the vacuum EMRI parameters. Here, in a general framework, we show that these correlations remain even if any modeled effect is absent in the actual signal, and that they cause inference bias when any effect in the signal is ignored 
in the analysis model. This worsens the overall measurability of the whole parameter set, challenging the constraints found by previous studies, and posing a general problem for the modeling and inference of beyond vacuum-GR effects in EMRIs. 
\end{abstract}

\maketitle 

\section{Introduction}\label{sec:intro}
    EMRIs are binary GW sources comprising a stellar mass compact object (CO) of mass $\mu$ orbiting a supermassive black hole (MBH) of mass $M$. Their parameters are expected to be constrained to sub-percent precision by the ESA-NASA observatory LISA~\citep{Amaro_Seoane_2015, Barack_2004,consortium2013gravitational,amaroseoane2017laser, Amaro_Seoane_2012,amaroseoane2012elisa}, enabling stringent tests of GR and the measurement of perturbative ``beyond vacuum-GR'' effects induced by modified gravity theories or astrophysical environments \cite{Amaro_Seoane_2007,Gair_2013,Babak_2017,Amaro_Seoane_2018}.
    
    EMRI waveforms are best modeled using black hole perturbation theory and self-force theory, where the Einstein field equations are solved by treating the gravitational field of the inspiralling CO as a perturbation to the Kerr spacetime of the central MBH \cite{Pound_2021, Barack_2018}. At leading order in the mass ratio $\mu/M$, this approach reduces approximately to a set of flux equations describing the long-term evolution of the system \cite{1973ApJ...185..635T, Hinderer_2008,
    Fujita_2020, Isoyama_2022, Hughes_2021}. A beyond vacuum-GR effect may then be introduced by directly solving the field equations with the system's stress-energy tensor altered by this effect, or by including the expected modifications due to the effect at the level of the fluxes (or the waveform itself). The latter approach is far more popular in the literature, due to its greater simplicity.
    
    In flux-based models, the flux equations modified by a general perturbative effect take the form
    \begin{align}
            \dot{E} &= \dot{E}_{\mathrm{GR}}(\boldsymbol{\psi})\left(1 + A(\alpha)\mathcal{F}(\boldsymbol{\psi},\boldsymbol{\phi})\right), \label{Edotcorrected}\\
    \dot{L} &= \dot{L}_{\mathrm{GR}}(\boldsymbol{\psi})\left(1 + B(\alpha)\mathcal{G}(\boldsymbol{\psi},\boldsymbol{\phi})\right), \label{Ldotcorrected}
    \end{align}
    where $\dot{E}_{\mathrm{GR}},\dot{L}_{\mathrm{GR}}$ are the leading-order model-dependent fluxes of the energy at infinity, $E$, and the axial component of angular momentum, $L$, respectively \cite{thorne2000gravitation}. The fluxes are typically calculated in the Teukolsky formalism \cite{Fujita_2020}\footnote{See, e.g., Eqs (1)--(3) in \cite{Isoyama_2022} or Eqs (3.26)--(3.28) in \cite{Hughes_2021}.}
    and depend only on the EMRI's vacuum-GR parameter vector $\boldsymbol{\psi}$, while $A\mathcal{F},B\mathcal{G}$ are the corrections induced by the perturbative effect.\footnote{The evolution $\dot{Q}$ of the Carter constant \cite{Carter:1968rr} (the third conserved quantity in Kerr orbits \cite{,Sago_2006}) is neglected here, since $\dot{E}, \dot{L}$ are more commonly modified in such treatments. Still, all of our arguments extend straightforwardly to the inclusion of $\dot{Q}$.} The amplitude of the effect is given by $A, B$, which in general can be functions of an underlying parameter $\alpha$ (such that $A, B \to 0$ as $\alpha \to 0$). The remaining functional dependence of the effect on any non-amplitude effect parameters $\boldsymbol{\phi}$ is contained in $\mathcal{F},\mathcal{G}$, which may also depend on $\boldsymbol{\psi}$. These modified fluxes are then used in GR waveform models $h(\dot{E},\dot{L})$ to generate beyond vacuum-GR waveforms (with the commonly made assumption that the effect does not modify the dependence of $h$ itself on $\dot{E},\dot{L}$, see for e.g.~\cite{Kocsis_2011} and \cite{Yunes_2011}).
    
    This general framework has been utilized by many studies to model perturbative effects in EMRIs. For example, Chamberlain et al. \cite{Chamberlain_2017} found that the effect of GW dipole radiation, predicted by some scalar-tensor modified gravity theories \cite{Blanchet2006, Will_2006, Brans:1961sx, will_1993} (which perturbs the vacuum-GR energy flux as $\dot{E} = \dot{E}_\mathrm{GR} \left(1+ Bp^{n_\mathrm{dip}}\right)$ \cite{Barausse_2016} with $p\equiv p(\boldsymbol{\psi}),B\equiv\alpha,n_\mathrm{dip}\equiv\boldsymbol{\phi}$), can be constrained by LISA to a precision several orders of magnitude better than current ground-based GW detectors.
    More recently, Speri et al. \cite{Speri_2023} found that the planetary migration torque of a thin accretion disk surrounding an EMRI (which modifies the angular momentum flux as $\dot{L}=\dot{L}_\mathrm{GR}\left(1 + A\left(p/10\right)^n\right)$ \cite{Barausse_2014, Kocsis_2011} with $p\equiv p(\boldsymbol{\psi}),A\equiv\alpha,n\equiv\boldsymbol{\phi}$) is measurable by LISA to $\sim 20\%$ precision. Many other examples that set similarly tight constraints on beyond vacuum-GR effects can be found in the literature, e.g., \cite{Yunes_2010,Kocsis_2011,Cardoso_2011, Yunes_2012,Barausse_2014,Barsanti_2023}.
    
    The use of flux corrections with the power-law form $Ap^n$ (where $p$ is the evolving dimensionless separation of the orbit) is especially prevalent in the literature. This form enables one to gauge the post-Newtonian (PN) order at which the effect becomes relevant (through $n$), and its strength with the vacuum-GR evolution (through $A$). Power-law models implicitly treat the effect in an orbit-averaged way, i.e., the manifestation of the effect on the timescale of orbital motion is neglected, and the accumulated effect is described on the longer radiation-reaction timescale as a secular ``drift'' from vacuum-GR orbital evolution. However, such modeling may strip the effect of any unique features, introducing correlations with the vacuum-GR parameters and other modeled effects, which may be simultaneously present in the realm of modified gravity theories \cite{Wang_2022} or environment-rich EMRIs \cite{Kocsis_2011}. 
    While most current studies set constraints on such effects in isolation, a more realistic analysis would thus have to account for these correlations.
    
\begin{figure}
     \centering
         \includegraphics[width=1.0\columnwidth]{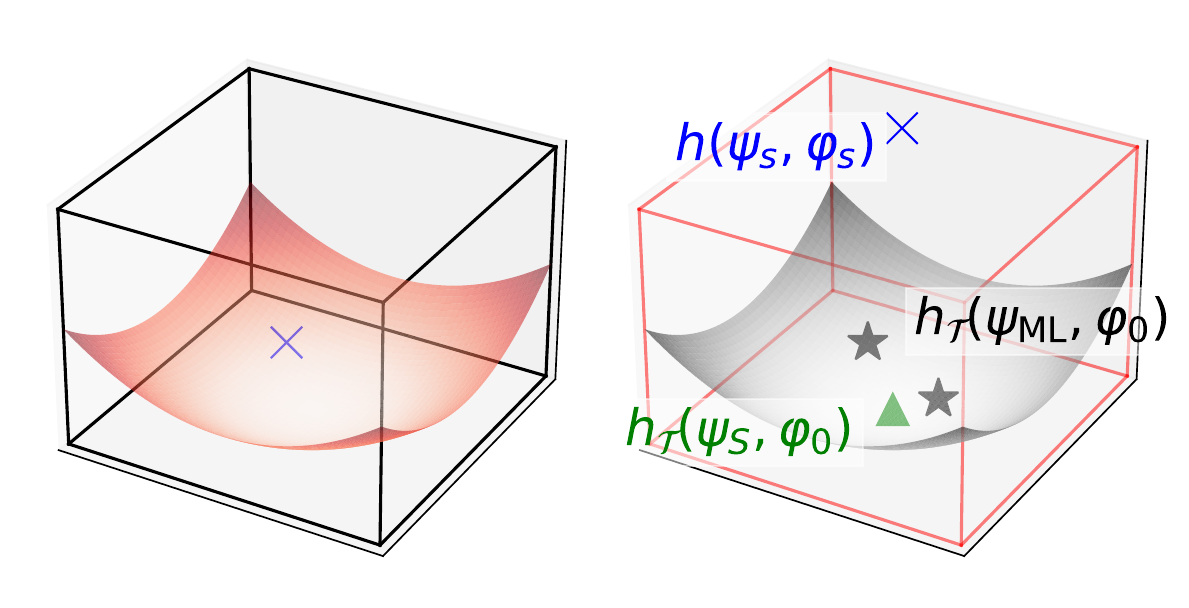}
     \caption{\justifying A schematic depiction of Case \rom{1} with the signal manifold $\mathcal{S}$ (red surface) embedded in the template manifold $\mathcal{T}$ (black box) (left) and Case \rom{2} with $\mathcal{T}$ (black surface) $\subset\mathcal{S}$ (red box) (right). The blue cross represents the true signal $s$, the black stars are the posterior modes, and the green triangle is the template that shares the signal parameters restricted to $\mathcal{T}$. 
     }
        \label{fig:templatevssignal}
\end{figure}

    Here, we develop a general mathematical framework to demonstrate 
    some of the main challenges posed by the joint analysis of multiple perturbative effects in EMRI modeling and inference. Through a Fisher information matrix (FIM) analysis of the standard GW parameter likelihood,\footnote{The FIM approximation is oft viewed as inferior to full likelihood sampling in GW analysis studies, but it is a perfectly viable tool for the analysis of local correlations if not higher-order likelihood moments, and its reliability is also much easier to ensure than that of sampling in the regime of strong correlations.} we study two main cases: Case \rom{1}, in which the signal $s\in\mathcal{S}$ (where $\mathcal{S}$ is the image space of a putative ``true'' model) also lies in the template manifold $\mathcal{T}$ (the image space of the analysis model); and Case \rom{2}, in which it does not.\footnote{In reality, detector noise ensures Case \rom{2}, but we neglect it here as its interaction with parameter inference is well understood.} Case \rom{1} has the subcases $\mathcal{S}\subseteq\mathcal{T}$ (e.g., a null test of GR; left panel of Fig.~\ref{fig:templatevssignal}) and $\mathcal{S}\not\subseteq\mathcal{T}$ (e.g., a mismodeled null test), while Case \rom{2} has the subcases $\mathcal{T}\subset\mathcal{S}$ (e.g., a neglected effect; right panel of Fig.~\ref{fig:templatevssignal}) and $\mathcal{T}\not\subset\mathcal{S}$ (e.g., a mismodeled effect).
    After establishing the generic cases, we illustrate our results with a taxonomy of representative examples, comment on their consequences for various themes in EMRI modeling and inference,
    and discuss possible ways to minimize their impact on the science output of EMRI observations with LISA.


\section{General results}\label{sec:analytic} 
To unify the treatment of Cases \rom{1} and \rom{2}, we define the unique waveform model $h(\boldsymbol{\theta})$ whose parameter set is the union of the signal and template parameter sets, and whose restrictions to those sets are the signal and template models $h_\mathcal{S}$ and $h_\mathcal{T}$ respectively. 
The GW likelihood corresponding to $h$ is $\mathcal{L}(\boldsymbol{\theta}|s)=\exp{-1/2\langle s-h(\boldsymbol{\theta})|s-h(\boldsymbol{\theta})\rangle}$ (where $\left<\cdot|\cdot\right>$ is the detector-noise-weighted inner product on the space of fixed-length time series \cite{Barack_2004}). The FIM of $\mathcal{L}$ is written in terms of waveform derivatives as \cite{Vallisneri_2008}
    \begin{align}
        \Gamma_{ij}(\boldsymbol{\theta}) =& \left.\left<\tfrac{\partial h}{\partial \theta_i}\right|\tfrac{\partial h}{\partial \theta_j}\right>\nonumber\\
        =& \left.\left<\tfrac{\partial h}{\partial \dot{E}}\tfrac{\partial\dot{E}}{\partial\theta_i} + \tfrac{\partial h}{\partial \dot{L}}\tfrac{\partial\dot{L}}{\partial\theta_i}\right|\tfrac{\partial h}{\partial \dot{E}}\tfrac{\partial\dot{E}}{\partial\theta_j} + \tfrac{\partial h}{\partial \dot{L}}\tfrac{\partial\dot{L}}{\partial\theta_j}\right>,\label{Gammachainrule}
    \end{align}
    and is evaluated at the maximum-likelihood estimate $\boldsymbol{\theta}_\mathrm{ML}$. The inverse FIM, $\Sigma=\Gamma^{-1}$, is then by definition the covariance matrix of the normal approximation to $\mathcal{L}$.
\subsection{Case \rom{1} results}\label{sec:analyticcase1}
    We first examine Case \rom{1} with $\mathcal{S}\subseteq\mathcal{T}$, such that $s=h(\boldsymbol{\theta}_\mathrm{ML})$. The subcase $\mathcal{S}\not\subseteq\mathcal{T}$ requires the restriction of all quantities to $\mathcal{T}$, but is otherwise identical. As a representative example, we consider a flux-based EMRI model modified by two beyond vacuum-GR effects:
\begin{align}
    \dot{E} &= \dot{E}_{\mathrm{GR}}(\boldsymbol{\psi})\left(1 + \Sigma_{k=1}^2A_k(\alpha_k)\mathcal{F}_k(\boldsymbol{\psi},\boldsymbol{\phi_k})\right), \label{Edottwoeffects}\\
    \dot{L} &= \dot{L}_{\mathrm{GR}}(\boldsymbol{\psi})\left(1 + \Sigma_{k=1}^2B_k(\alpha_k)\mathcal{G}_k(\boldsymbol{\psi},\boldsymbol{\phi_k})\right), \label{Ldottwoeffects}
\end{align}
which yields the corresponding FIM $\Gamma(\boldsymbol{\psi},\alpha_1,\boldsymbol{\phi}_1,\alpha_2,\boldsymbol{\phi}_2)$. Case \rom{1} is then obtained by letting one of the effects, say $k=2$, vanish in the signal. While the FIM is singular at $\alpha_2 = 0$, in the limit $\alpha_2 \to 0$ in $s$, 
the rows and columns of the FIM corresponding to the components of $\boldsymbol{\phi}_2$ will vanish in proportion to $\alpha_2$, making $\boldsymbol{\phi}_2$ unmeasurable. 
Equivalently, the precisions $(\Sigma_{\boldsymbol{\phi}_2\boldsymbol{\phi}_2})^{1/2}$ to which $\boldsymbol{\phi}_2$ can be measured blow up as $\alpha_2 \to 0$. The implied presence of an unmeasurable parameter set $\boldsymbol{\phi}_2$ in the analysis model will make the inference of the full parameter space inefficient. The setup also highlights a more important consequence, i.e. all correlation coefficients $\rho_{ij} \equiv \Sigma_{ij}/(\Sigma_{ii}\Sigma_{jj})^{1/2}$ are $\mathcal{O}(1)$ as one of the effects vanishes ($\alpha_2 \to 0$, without loss of generality), as we show in Appendix~\ref{app:correlations}.

Such finite correlations between the otherwise well-constrained parameter set (including the vacuum-GR and other perturbative effects' parameters) with $\boldsymbol{\phi}_2$ 
near $\alpha_2 = 0$ imply that constraints on them will also generally degrade in proportion to these correlations. 
This holds even when $A_2,B_2$ (and their derivatives $A_2',B_2'$, if applicable) vanish at different rates, which can represent a broad class of beyond vacuum-GR models.

\subsection{Case \rom{2} results}
Correlations between parameters also significantly impact inference in Case \rom{2}, where the signal contains an effect that is neglected or mismodeled. We restrict to the subcase $\mathcal{T}\subset\mathcal{S}$ for simplicity,\footnote{The subcase $\mathcal{T}\not\subset\mathcal{S}$ requires more bookkeeping but is straightforward to extend, and will be addressed in follow-up work.} and partition the parameters $\boldsymbol{\theta}$ more generally as $(\boldsymbol{\psi},\boldsymbol{\varphi})$, where the $\boldsymbol{\varphi}$ describe an effect that vanishes for some values $\boldsymbol{\varphi}_0$. The signal is $s=h(\boldsymbol{\psi}_s,\boldsymbol{\varphi}_s)\not\in\mathcal{T}$, for some $\boldsymbol{\varphi}_s\neq\boldsymbol{\varphi}_0$. Note that the analysis waveform model is now not $h\equiv h_\mathcal{S}$ but its restriction to $\mathcal{T}$, i.e., $h_\mathcal{T}(\boldsymbol{\psi})\equiv h_\mathcal{S}(\boldsymbol{\psi},\boldsymbol{\varphi}=\boldsymbol{\varphi}_0)$. Its corresponding likelihood $\mathcal{L}_\mathcal{T}$ yields the maximum-likelihood estimate $(\boldsymbol{\psi}_\mathrm{ML},\boldsymbol{\varphi}_0)$. The FIM on $\mathcal{T}$ is then the submatrix $\Gamma_{\boldsymbol{\psi}\boldsymbol{\psi}}$ of the ``FIM'' $\Gamma$ on $\mathcal{S}$, evaluated at $(\boldsymbol{\psi}_\mathrm{ML},\boldsymbol{\varphi}_0)$.

Cutler and Vallisneri \cite{Cutler_2007} showed that the inference bias $\Delta\boldsymbol{\psi}\equiv\boldsymbol{\psi}_\mathrm{ML}-\boldsymbol{\psi}_s$ is given to leading order by
\begin{equation}\label{eq:cv}
    \Delta\boldsymbol{\psi}=(\Gamma_{\boldsymbol{\psi}\boldsymbol{\psi}})^{-1}\cdot\langle(\partial_{\boldsymbol{\psi}} h_\mathcal{T})(\boldsymbol{\psi}_\mathrm{ML})|s-h_\mathcal{T}(\boldsymbol{\psi}_s)\rangle,
\end{equation}
in our notation. The original application of Eq.~\eqref{eq:cv} is to estimate $\boldsymbol{\psi}_s$ given $\boldsymbol{\psi}_\mathrm{ML}$, and thus relies on approximating the waveform difference $s-h_\mathcal{T}(\boldsymbol{\psi}_s)$ as $h_\mathcal{S}(\boldsymbol{\psi}_\mathrm{ML})-h_\mathcal{T}(\boldsymbol{\psi}_\mathrm{ML})$. This approach breaks down here since $\mathcal{T}\subset\mathcal{S}$, but is not actually needed since we assume that $\boldsymbol{\psi}_s$ is known instead (and wish to estimate the bias $\Delta\boldsymbol{\psi}$). We simply use the fact that to leading order, both the waveform derivatives and the FIM are invariant as $(\boldsymbol{\psi}_\mathrm{ML},\boldsymbol{\varphi}_0)\to(\boldsymbol{\psi}_s,\boldsymbol{\varphi}_0)$ (the whole approach is invalid anyway if one is not in the linear-signal regime).

We now reformulate the Cutler-Vallisneri approach for the $\mathcal{T} \subset \mathcal{S}$ case. By definition, $\boldsymbol{\psi}_\mathrm{ML}$ satisfies
\begin{equation}\label{eq:mlconstraint}
    \langle\partial_{(\boldsymbol{\psi},\boldsymbol{\varphi})} h_\mathcal{S}(\boldsymbol{\psi},\boldsymbol{\varphi}=\boldsymbol{\varphi}_0)|_{\boldsymbol{\psi}=\boldsymbol{\psi}_\mathrm{ML}}|s-h_\mathcal{S}(\boldsymbol{\psi}_\mathrm{ML},\boldsymbol{\varphi}_0)\rangle=0.
\end{equation}
Note that the derivative vector reduces to $((\partial_{\boldsymbol{\psi}} h_\mathcal{T})(\boldsymbol{\psi}_\mathrm{ML}),0)$, and is \emph{not} generally equivalent to $(\partial_{(\boldsymbol{\psi},\boldsymbol{\varphi})} h_\mathcal{S})(\boldsymbol{\psi}_\mathrm{ML},\boldsymbol{\varphi}_0)$, since the derivative of an effect amplitude parameter does not generally vanish at $\boldsymbol{\varphi}_0$.

One may nonetheless proceed with the Cutler--Vallisneri derivation \cite{Cutler_2007} on the ambient manifold $\mathcal{S}$ rather than $\mathcal{T}$. Expanding $h_\mathcal{S}(\boldsymbol{\psi}_s,\boldsymbol{\varphi}_0)$ about $(\boldsymbol{\psi}_\mathrm{ML},\boldsymbol{\varphi}_0)$ to first order in $(\Delta\boldsymbol{\psi},0)$, and substituting into the maximum-likelihood constraint \eqref{eq:mlconstraint}, we find
\begin{widetext}
\begin{align}
    \langle\partial_{(\boldsymbol{\psi},\boldsymbol{\varphi})} h_\mathcal{S}(\boldsymbol{\psi},\boldsymbol{\varphi}=\boldsymbol{\varphi}_0)|_{\boldsymbol{\psi}=\boldsymbol{\psi}_\mathrm{ML}}|s-h_\mathcal{S}(\boldsymbol{\psi}_s,\boldsymbol{\varphi}_0)\rangle=\langle\partial_{(\boldsymbol{\psi},\boldsymbol{\varphi})} h_\mathcal{S}(\boldsymbol{\psi},\boldsymbol{\varphi}=\boldsymbol{\varphi}_0)|_{\boldsymbol{\psi}=\boldsymbol{\psi}_\mathrm{ML}}|(\partial_{(\boldsymbol{\psi},\boldsymbol{\varphi})} h_\mathcal{S})(\boldsymbol{\psi}_\mathrm{ML},\boldsymbol{\varphi}_0)\rangle\cdot(\Delta\boldsymbol{\psi},0)\nonumber\\
    \Rightarrow
    \begin{bmatrix}
        \langle\partial_{\boldsymbol{\psi}} h_\mathcal{S}(\boldsymbol{\psi},\boldsymbol{\varphi}=\boldsymbol{\varphi}_0)|_{\boldsymbol{\psi}=\boldsymbol{\psi}_\mathrm{ML}}|s-h_\mathcal{S}(\boldsymbol{\psi}_s,\boldsymbol{\varphi}_0)\rangle\\
        0
    \end{bmatrix}=
    \begin{bmatrix}
        \Gamma_{\boldsymbol{\psi}\boldsymbol{\psi}}(\boldsymbol{\psi}_\mathrm{ML},\boldsymbol{\varphi}_0)&\Gamma_{\boldsymbol{\psi}\boldsymbol{\varphi}}(\boldsymbol{\psi}_\mathrm{ML},\boldsymbol{\varphi}_0)\\
        0&0
    \end{bmatrix}\cdot
    \begin{bmatrix}
        \Delta\boldsymbol{\psi}\\
        0
    \end{bmatrix}.\label{eq:cvdiffdim}
\end{align}
\end{widetext}
Once again, note that the outer product on the right side is \emph{not} the FIM on $\mathcal{S}$, and in particular is non-invertible since it has zero rows. However, as both sides of \eqref{eq:cvdiffdim} have the same zero rows (corresponding to $\boldsymbol{\varphi}$), it simplifies to 
\begin{equation}\label{eq:cvapp}
    \langle(\partial_{\boldsymbol{\psi}} h_\mathcal{T})(\boldsymbol{\psi}_\mathrm{ML})|s-h_\mathcal{T}(\boldsymbol{\psi}_s)\rangle=\Gamma_{\boldsymbol{\psi}\boldsymbol{\psi}}(\boldsymbol{\psi}_\mathrm{ML})\cdot\Delta\boldsymbol{\psi},
\end{equation}
which is the Cutler--Vallisneri equation on $\mathcal{T}$, as expected.

Although not immediately evident in Eq.~\eqref{eq:cvapp}, the correlations between $\boldsymbol{\psi}$ and $\boldsymbol{\varphi}$ still affect the bias, but through the signal $s$ rather than the FIM on $\mathcal{T}$ (which does not depend on $\boldsymbol{\varphi}$ in the first place). We perform further expansions of $s=h_\mathcal{S}(\boldsymbol{\psi}_s,\boldsymbol{\varphi}_s)$ and $h_\mathcal{S}(\boldsymbol{\psi}_s,\boldsymbol{\varphi}_0)$ about $(\boldsymbol{\psi}_\mathrm{ML},\boldsymbol{\varphi}_0)$, and a substitution into Eq.~\eqref{eq:cvdiffdim} gives
\begin{equation}
    \begin{bmatrix}
        \Gamma_{\boldsymbol{\psi}\boldsymbol{\psi}}&\Gamma_{\boldsymbol{\psi}\boldsymbol{\varphi}}\\
        0&0
    \end{bmatrix}\cdot
    \begin{bmatrix}
        0\\
        \boldsymbol{\varphi}_s-\boldsymbol{\varphi}_0
    \end{bmatrix}=
    \begin{bmatrix}
        \Gamma_{\boldsymbol{\psi}\boldsymbol{\psi}}&\Gamma_{\boldsymbol{\psi}\boldsymbol{\varphi}}\\
        0&0
    \end{bmatrix}\cdot
    \begin{bmatrix}
        \Delta\boldsymbol{\psi}\\
        0
    \end{bmatrix},
\end{equation}
where $\Gamma_{\boldsymbol{\psi}\boldsymbol{\psi}}$ and $\Gamma_{\boldsymbol{\psi}\boldsymbol{\varphi}}$ are submatrices of the ``FIM'' on $\mathcal{S}$, evaluated at $(\boldsymbol{\psi}_s,\boldsymbol{\varphi}_0)$ instead of $(\boldsymbol{\psi}_\mathrm{ML},\boldsymbol{\varphi}_0)$. Thus
\begin{equation}\label{eq:cvnested}
    \Delta\boldsymbol{\psi}=(\Gamma_{\boldsymbol{\psi}\boldsymbol{\psi}})^{-1}\Gamma_{\boldsymbol{\psi}\boldsymbol{\varphi}}\cdot(\boldsymbol{\varphi}_s-\boldsymbol{\varphi}_0),
\end{equation}
which describes to leading order the bias incurred when performing inference with a template model that is a nested submodel of the signal model.

Via the matrix inversion lemma \cite{boyd2004convex}, we may also write the matrix $(\Gamma_{\boldsymbol{\psi}\boldsymbol{\psi}})^{-1}\Gamma_{\boldsymbol{\psi}\boldsymbol{\varphi}}$ in terms of submatrices of the inverse FIM $\Sigma=\Gamma^{-1}$:
\begin{align}\label{eq:corrbias}
    (\Gamma_{\boldsymbol{\psi}\boldsymbol{\psi}})^{-1}\Gamma_{\boldsymbol{\psi}\boldsymbol{\varphi}}=-\Sigma_{\boldsymbol{\psi}\boldsymbol{\varphi}}(\Sigma_{\boldsymbol{\varphi}\boldsymbol{\varphi}})^{-1},
\end{align}
which shows the explicit relationship between the parameter biases $\Delta\boldsymbol{\psi}$ and the parameter covariances $\Sigma_{\boldsymbol{\psi}\boldsymbol{\phi}}$. In obtaining Eqs~\eqref{eq:cvnested} and \eqref{eq:corrbias}, we have thus explicitly highlighted the dependence of induced biases on correlations between $\boldsymbol{\psi}$ and $\boldsymbol{\varphi}$ (through $\Sigma_{\boldsymbol{\psi}\boldsymbol{\varphi}}$), and on the value $\boldsymbol{\varphi}_s$ of the neglected parameter in the signal.


\section{Examples}
We now consider a representative population of signals from specific but typical models for vacuum-GR EMRIs and beyond vacuum-GR effects, to illustrate the quantitative impact of these parameter correlations. The vacuum-GR model we use describes the adiabatic evolution of circular and equatorial EMRIs in Kerr spacetime \cite{Hughes_2021}, as implemented in the \texttt{FastEMRIWaveforms} package
\cite{Chua:2020stf,PhysRevD.104.064047,Speri:2023jte}. While eccentric and inclined Kerr models are already available \cite{Isoyama_2022,Chua:2017ujo}, the current literature lacks models of environmental and modified gravity effects for such generic Kerr orbits. Extensions to these orbits may be important to fully gauge the implications of our analysis, and are left to future studies.

\subsection{Signal and population models}
As a representative set of signals, we consider $100$ sources with redshifted primary masses $M \in [10^6,10^7] M_\odot$ distributed according to the EMRI population model \texttt{M1} of Babak et al.~\cite{Babak_2017}, and uniformly distributed initial separations $p_0$ such that all sources are close to plunge after evolving for $T \in [3.0,4.0]$ years in the LISA band.
The other signal parameters have a smaller impact on the spread of our results, and are set to the same values for all sources: redshifted secondary mass $\mu = 50 M_\odot$; dimensionless MBH spin $a=0.9M$; initial azimuthal phase $\Phi_0=3.0$; sky location $\theta_S= \phi_S=0.2$; spin orientation $\theta_K = \phi_K=0.8$; and luminosity distance $d_L=1.5$ Gpc.
We further simplify the analysis by fixing the extrinsic parameters $\{\theta_S,\phi_S,\theta_K,\phi_K,d_L\}$ in our templates to the signal values. The vacuum-GR parameters that we infer are thus $\boldsymbol{\psi} \equiv (M,\mu,a,p_0,\Phi_0)$. All sources have signal-to-noise ratios $>20$ in the long-wavelength LISA response approximation \cite{PhysRevD.57.7089} with a zero noise realisation.

In addition, we consider two beyond vacuum-GR effect models in this work. First is the planetary migration (PM) effect, which can be modeled as a perturbation to $\dot{L}$ as $ A_{\mathrm{PM}}\left(p/10M\right)^{n_{\mathrm{PM}}}$ \cite{Kocsis_2011,Speri_2023}. In Shakura--Sunyaev $\alpha$ and $\beta$ discs in the geometrically thin, optically thick configuration, PM is expected to significantly alter the evolution of EMRIs. To comment on the effectiveness of descriptive modeling, 
we also examine the impact of restoring a factor of $F_m = (1-\sqrt{p_\mathrm{in}/p})^{m/4}$ to the PM effect (where $p_{\mathrm{in}}$ is the assumed inner edge of the disk and $m$ is a free parameter)~\cite{Barausse_2014}. The second effect is the time-varying gravitational constant (GC), with expected tight constraints from LISA EMRIs \cite{Chamberlain_2017, Yunes_2010}. The GC effect also scales as a negative-PN order effect, perturbatively added to $\dot{L}_\mathrm{GR}$ as $A_{\mathrm{GC}}(p/M)^{n_{\mathrm{GC}}}$~\cite{Speri_2023}. The net torque of an EMRI perturbed by the PM and GC effects is thus
\begin{align}\label{eq:jointtorque}
    \dot{L} = \dot{L}_{\mathrm{GR}}(\boldsymbol{\psi})\left(1 + A_{\mathrm{PM}}\left(\tfrac{p}{10M}\right)^{n_{\mathrm{PM}}}F_m + A_{\mathrm{GC}}\left(\tfrac{p}{M}\right)^{n_{\mathrm{GC}}}\right),
\end{align}
such that $(\alpha_1,\boldsymbol{\phi}_1,\alpha_2,\boldsymbol{\phi}_2) \equiv (A_\mathrm{PM},(n_\mathrm{PM},m), A_\mathrm{GC},n_\mathrm{GC})$.

\begin{figure}
\centering
\begin{subfigure}{0.42\textwidth}
\includegraphics[width=\textwidth]{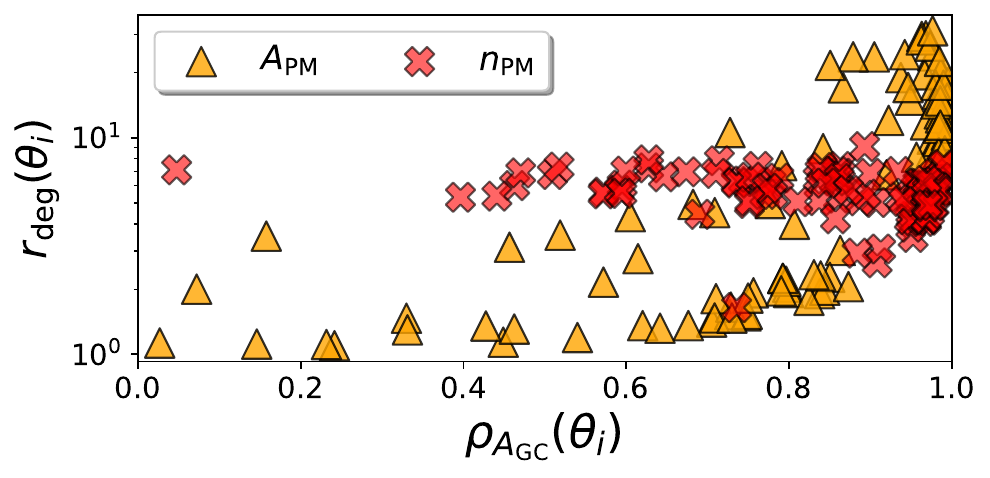}
\end{subfigure}

\begin{subfigure}{0.42\textwidth}
    \includegraphics[width=\textwidth]{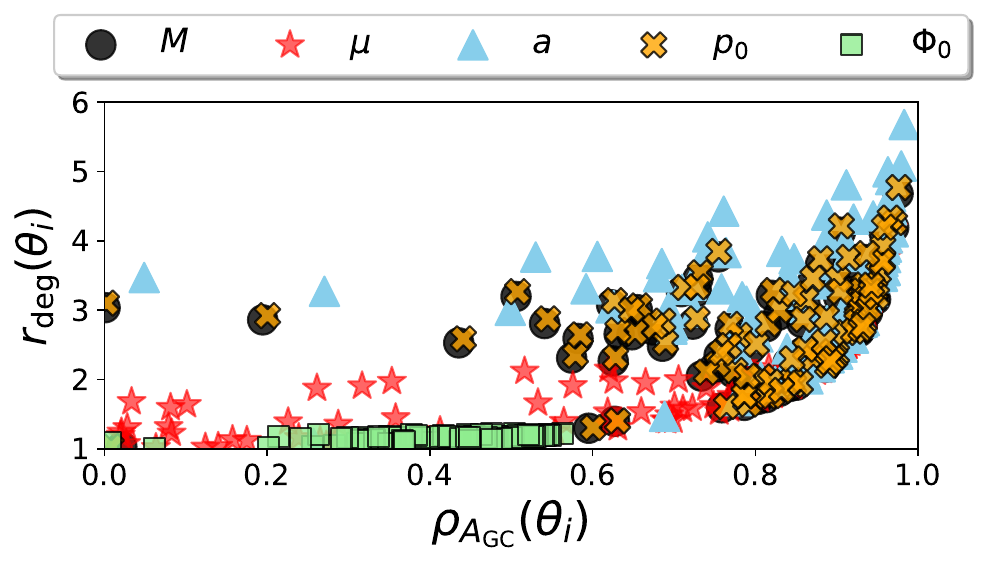}
\end{subfigure}

\caption{\justifying Relation between the degradation ratio $r_\mathrm{deg}(\theta_i)$ for the source parameters $\theta_i$ of 100 representative EMRI signals, and their correlations $\rho_{A_\mathrm{GC}}(\theta_i)$ with the amplitude of the GC effect (vanishing in the signal and unmeasurable in the analysis). Top panel: PM parameters; bottom panel: vacuum-GR parameters.}
\label{fig:Examples}
\end{figure}

\subsection{Case \rom{1} examples} We first examine the joint analysis of the PM and GC effects in signals that only have a significant PM effect. 
Assuming an $\alpha$-disk model in the sensitivity range of LISA \cite{Speri_2023}, the PM parameters of the signal are set as $(A_{\mathrm{PM}}, n_{\mathrm{PM}},m) = (1.92M_6\times10^{-5}
,8,0)$, where $M_6 \equiv M/(10^6M_\odot)$, 
while the (unmeasurable) GC parameters are set as $(A_{\mathrm{GC}}, n_{\mathrm{GC}}) = (10^{-12},4)$. The degradation of measurement precision for parameter $\theta_i$ in the joint analysis, relative to an analysis with a PM-only template, can be quantified by the ratio of its standard deviations in the two analyses: $r_\mathrm{deg}(\theta_i) \equiv \sigma_{\rm joint}(\theta_i)/\sigma_{\rm PM}(\theta_i)$, where $\sigma_*(\theta_i)\equiv(\Sigma_{ii})^{1/2}$ in each analysis. As discussed earlier, this degradation ratio is expected to increase for a given $\theta_i$ if it has a higher correlation $\rho_{A_\mathrm{GC}}(\theta_i)\equiv\rho_{iA_\mathrm{GC}}$ with the unmeasurable amplitude parameter $A_\mathrm{GC}$.

This trend is observed in Fig.~\ref{fig:Examples}, which plots $r_\mathrm{deg}$ against $\rho_{A_\mathrm{GC}}$ for the PM parameters (top panel) and the vacuum-GR parameters (bottom panel) of our 100 representative signals. The measurability of the PM parameters is highly degraded due to their generally stronger correlations with $A_{\rm GC}$, with $r_\mathrm{deg}(A_\mathrm{PM})>10$ in over 30\% of the signals and $r_\mathrm{deg}(n_\mathrm{PM})>6$ in over 40\% of the signals. Some of the vacuum-GR parameters are also degraded significantly, with $r_\mathrm{deg}>3$ for each of $\{M,a,p_0\}$ in around 40--60\% of the signals.
Similar results are obtained even if $n_{\rm GC}$ is fixed in the template, as routinely done in modified gravity studies (e.g., \cite{Chamberlain_2017})
However, restoring the $F_m$ factor to the PM effect (and adding $m = -12$ to the parameter set) reduces the parameter correlations as expected, such that $r_\mathrm{deg}$ is now $<3$ for all PM and vacuum-GR parameters in around 90\% of the signals.

\subsection{Case \rom{2} examples} 
\begin{figure}
\includegraphics[width=\columnwidth]{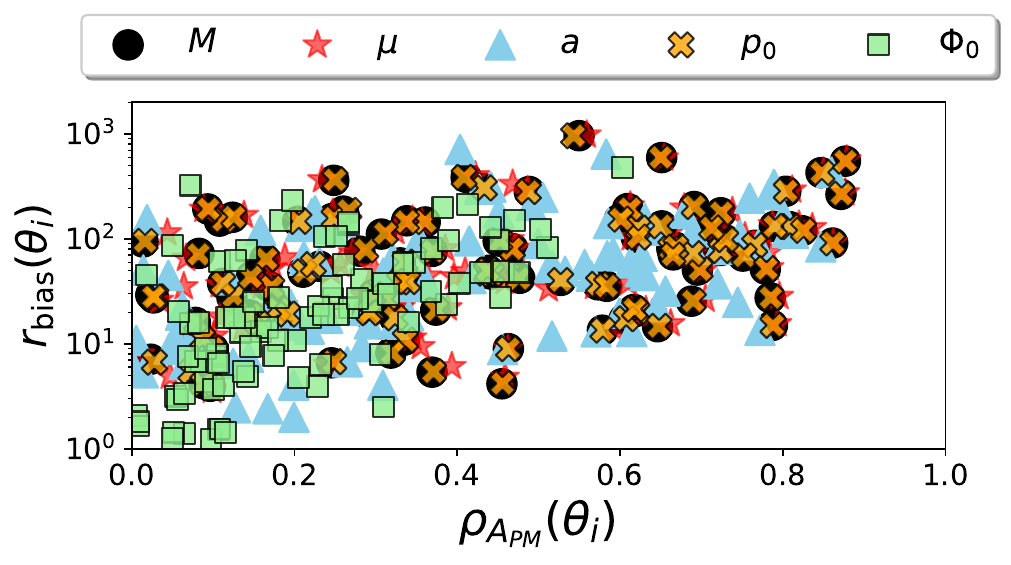}
\caption{\justifying Relation between the bias-to-precision ratio $r_\mathrm{bias}(\theta_i)$ for the vacuum-GR parameters $\theta_i$ of 100 representative EMRI signals, and their correlations $\rho_{A_\mathrm{PM}}(\theta_i)$ with the amplitude of the PM effect (present in the signal but neglected in the analysis).}
\label{fig:case2}
\end{figure}

We now consider the analysis of PM-affected EMRI signals using vacuum-GR templates. As in the Case I examples, the PM parameters are set as $(A_{\mathrm{PM}}, n_{\mathrm{PM}},m) = (1.92M_6\times10^{-5},8,0)$. Using the leading-order bias estimate \eqref{eq:cvnested} with $\boldsymbol{\varphi}_s=(1.92M_6\times10^{-5},8)$ and $\boldsymbol{\varphi}_0=(0,0)$,\footnote{For a power-law effect, $n_{\mathrm{PM}}$ is a redundant degree of freedom in $\boldsymbol{\varphi}_0$ since $A_{\mathrm{PM}}=0$, but the bias estimate does not depend on any choice of $n_{\mathrm{PM}}$ since the corresponding columns of $\Gamma_{\boldsymbol{\psi}\boldsymbol{\varphi}}$ are zero.} we find that the bias-to-precision ratio $r_\mathrm{bias}(\theta_i)\equiv|\Delta\theta_i|/\sigma_\mathrm{GR}(\theta_i)$ for all vacuum-GR parameters $\boldsymbol{\psi}$ is $>10$ for around 80\% of the analyzed signals, as shown in Fig~\ref{fig:case2}. Thus, a considerable fraction of the analyses incur significant ($10\sigma$ or higher) biases. Also, $r_\mathrm{bias}$ tends to be higher for a given parameter if it is more correlated with the neglected amplitude parameter $A_\mathrm{PM}$; this dependence is explicitly evident from Eq.~\eqref{eq:cvnested}.

Along with biases, note that Case \rom{2} generically leads to multimodalities in inference, with the posterior modes corresponding to any template whose difference from $s$ is orthogonal to $\mathcal{T}$ \cite{Cutler_2007,Chua:2018afi,Chua_2022} (as depicted in Fig.~\ref{fig:templatevssignal}). Beyond its inconveniencing of posterior sampling in GW inference, this phenomenon severely hinders the prospects of searching for EMRIs in LISA data \cite{Chua_2022}. In the context of a joint analysis of multiple perturbative effects, the degree of multimodality is likely to increase with the number of effects under consideration (both modeled and unmodeled). However, the linear-signal framework we adopt here describes only local correlations, and is thus unsuitable for a study of multimodality.

\section{Discussion}
\subsection{Consequences for EMRI science}
\subsubsection{The ppE formalism}
We now highlight how our analysis extends to other current themes in the modeling and analysis of LISA EMRIs. The parametrized post-Einsteinian (ppE) formalism~\cite{Yunes_2009} 
is a popular framework for modeling modified gravity effects at the level of the waveform. When constraining multiple such effects simultaneously, the modified template would (by natural extension) take the form
\begin{align}\label{eq:ppE}
    &\tilde{h}(f) = \tilde{h}_{\mathrm{GR}}(f)\left(1 + \Sigma_kA_ku^{a_k}\right)\exp[i\left(\Sigma_kB_ku^{b_k}\right)]
\end{align}
in our notation, where $\tilde{h}_\mathrm{GR}(f)$ is the vacuum-GR waveform with $u \equiv f\pi(M\mu)^{3/5}/(M+\mu)^{1/5}$ and $\{A_k,B_k,a_k,b_k\}$ the $k$-th effect's ppE parameters.

With the formalism developed in Section \ref{sec:analyticcase1}, we now show that an effect modeled by the ppE parameters will have non-zero (and potentially large) correlations with all other parameters, despite it being absent in the true signal. We focus on the $k=2$ effect vanishing in the signal, such that the corresponding amplitudes $A_2, B_2\to0$. The partial derivatives of $\tilde{h}(f)$ with respect to $(a_2, b_2)$ also vanish proportionally to $(A_2, B_2)$:
\begin{align}
    &\frac{\partial\tilde{h}(f)}{\partial a_2} = A_2 \times u^{a_2} \mathcal{H}, \label{dela2}\\
    &\frac{\partial\tilde{h}(f)}{\partial b_2} = B_2 \times iu^{b_2} (1+A_1u^{a_1}+A_2u^{a_2}) \mathcal{H},\label{delb2}
\end{align}
where
\begin{align}
    &\mathcal{H} = \tilde{h}_\mathrm{GR}(f) \ln{(u)}\exp[i\left(B_1u^{b_1}+B_2u^{b_2}\right)].
\end{align}

Since the Fourier transform commutes with partial derivatives ($\tilde{(\partial h)}=\partial\tilde{h}$) in the standard matched-filtering inner product \cite{Barack_2004} (weighted by detector noise with power spectral density $S_n(f)$), we have
\begin{align}
    \Gamma_{ij} = \left.\left<\frac{\partial h}{\partial\theta_i}\right|\frac{\partial h}{\partial\theta_j}\right> = 2\int_0^{\infty} \frac{(\partial_i\tilde{h})^\dagger\partial_j\tilde{h} + (\partial_j\tilde{h})^\dagger\partial_i\tilde{h}}{S_n(f)}\,df.\label{Fisher}
\end{align}
It is then straightforward to see that $A_2$ factors out of every row and column of the FIM that involves $a_2$, while $B_2$ factors out of every row and column that involves $b_2$. Thus, similar to the argument in Section~\ref{sec:analyticcase1} and Appendix~\ref{app:correlations}, the correlations between $a_2$ or $b_2$ and all other parameters are $\mathcal{O}(1)$ as the $k=2$ effect vanishes. At the same time, the precisions to which $a_2,b_2$ can be measured blow up in the same limit, and so the inference of other parameters is generally worsened as well, which can severely hinder joint inference. 

Even if $a_2, b_2$ are held fixed in the template, the correlations may still be large, owing to the mathematically similar form of the modeled perturbations. Separately, depending on the correlations between the parameters, a generic GR-consistent effect (like PM) may show up as a finite ppE parameter in the inference of Eq.~\eqref{eq:ppE}, giving rise to false alarms of deviations from GR (as also discussed in the context of flux-level modifications in Speri et al. \cite{Speri_2023}). The formalism developed in this paper can be readily employed to study the severity of such generic false alarms on tests of GR (see \cite{Gupta:2024gun} for a review).

\subsubsection{The EMRI secondary spin}
To meet the science requirements of LISA, the long-term evolution of vacuum-GR EMRIs must be described accurately up to first post-adiabatic (1PA) order \cite{Hinderer_2008,Pound_2021} when expanding in the small mass ratio $\varepsilon\equiv\mu/M$, i.e., models must include all $\mathcal{O}(\varepsilon)$ corrections to the leading-order adiabatic (0PA) evolution \cite{Hughes_2021, Isoyama_2022}. The spin $\chi$ of the secondary mass becomes relevant at this order \cite{Mathews_2022}. For a circular and equatorial EMRI with a spinning secondary, the energy flux model can be written as \cite{burke2023accuracy}
\begin{equation}\label{eq:secspin}
    \dot{E} = \dot{E}_\mathrm{0PA}(\boldsymbol{\psi}_0) + \varepsilon\dot{E}_\mathrm{SF}(\boldsymbol{\psi}_0) + \varepsilon\dot{E}_{\chi}(\boldsymbol{\psi}_0,\chi)+\mathcal{O}(\varepsilon^2),
\end{equation}
where $\boldsymbol{\psi}_0$ is the 0PA vacuum-GR parameter set, 
$\dot{E}_\mathrm{0PA}$ is the adiabatic flux, and the 1PA term has been split into two: $\dot{E}_{\chi}$, which contains all contributions from $\chi$, and $\dot{E}_\mathrm{SF}$, which contains all other self-force corrections.

Eq.~\eqref{eq:secspin} can be cast in the form of Eq.~\eqref{Edotcorrected}, with $\dot{E}_\mathrm{GR}\equiv\dot{E}_\mathrm{0PA}+\varepsilon\dot{E}_\mathrm{SF}$, $\mathcal{F}\equiv\dot{E}_\chi/(\dot{E}_\mathrm{0PA}+\varepsilon\dot{E}_\mathrm{SF})$, and $(\boldsymbol{\psi},A,\boldsymbol{\phi})\equiv(\boldsymbol{\psi}_0,\varepsilon,\chi)$. Since $\varepsilon \ll 1$ for an EMRI, $\chi$ is generally poorly constrained, if at all measurable (in the same way that $(\Sigma_{\boldsymbol{\phi}\boldsymbol{\phi}})^{1/2}\to\infty$ as $A \to 0$). This was observed by Burke et al. \cite{burke2023accuracy}, who found that $\chi$ in a circular Schwarzschild EMRI 
is significantly better measured when $\varepsilon = 10^{-4}$ than when $\varepsilon = 10^{-5}$. Previous studies also drew similar conclusions on the overall measurability of $\chi$ \cite{Piovano:2021iwv, Huerta_2011}. At the same time, 
the correlations between $\chi$ and $\boldsymbol{\psi}_0$ do not vanish in general, which would degrade the inference precision of the latter relative to using a template model with no secondary spin (this was not examined in \cite{burke2023accuracy}).



\subsection{Directions forward}

Strong correlations among beyond vacuum-GR effects in EMRIs typically arise due to simplistic models of those effects. The most direct way forward is thus to improve the modeling end, e.g., by adding detail to the description of the PM effect as shown earlier, or as attempted in \cite{cole2022disks}. Note that improved modeling of such effects still does not guarantee their decoupling in inference, simply because many of them are inherently perturbative (small-amplitude) and secular (long-timescale) in nature; see, e.g., \cite{Maselli_2020, Polcar_2022}. In contrast, if an effect manifests as a transient resonance \cite{Hinderer_2008,Berry_2016,Gupta:2022fbe,mukherjee2020resonant,Speri:2021psr}, it is likely to be more distinguishable from secular-type effects.

If more detailed models are not available (be it due to modeling difficulty or the actual simplicity of the effect), there are limited solutions on the analysis end. In similar spirit to ``agnostic'' parametric tests of GR, one could construct a beyond vacuum-GR model with only a single set of effect parameters, to perform a null test of the vacuum-GR hypothesis and to measure any potential deviation (e.g., the ppE formalism, or parametrised tests of GR used in ground-based observing \cite{Li_2012,Saleem_2022,Krishnendu:2021fga}). The functional dependence of the model on these parameters would attempt to generically describe most perturbative deviations (e.g., a power law, or perturbations to PN coefficients). In the context of the LISA global fit strategy \cite{Cornish_2005,Vallisneri_2009,Babak_2010}, this scheme would be more practically viable than the additive modeling of multiple effects; however, it would be subject to the usual pitfalls when introducing arbitrary degrees of freedom into a physical model \cite{Chua:2020oxn}. For example, any such model will be far more sensitive to some effects than others, potentially leading to severe selection bias. Furthermore, if multiple effects are truly present in the signal, the approach hinges strongly on our ability to find and interpret the multiple posterior modes.

An alternative strategy 
could be to rely on population-level studies. A catalog of detected sources might provide combined evidence for the presence of global effects (e.g., the time-varying gravitational constant) over local ones (e.g., accretion disks), thus decoupling them within a hierarchical Bayesian framework (e.g., \cite{Mandel:2018mve}). Such approaches and others will be needed to circumvent the problems caused by effect correlations and to unlock the full potential of EMRIs as high-precision probes of modified gravity and astrophysical environments.


\begin{acknowledgments}
SK thanks Enrico Barausse and Laura Sberna for insightful discussions. AJKC thanks Nicolas Yunes, Enrico Barausse, and other participants of the Asymmetric Binaries Meet Fundamental Astrophysics workshop at GSSI for helpful interactions. SK acknowledges the computational resource accessed from NUS IT Research Computing group and the support of the NUS Research Scholarship (NUSRS).
\end{acknowledgments}

\appendix

\section{Parameter correlations for a beyond vacuum-GR EMRI do not vanish if an effect is absent}\label{app:correlations}

We show here that all correlation coefficients $\rho_{ij} \equiv \Sigma_{ij}/(\Sigma_{ii}\Sigma_{jj})^{1/2}$ are $\mathcal{O}(1)$ as one of the effects vanishes ($\alpha_2 \to 0$, without loss of generality). For $\alpha_2$ and each component of $\boldsymbol{\phi}_2$, we have for all other parameters $\theta$:
\begin{widetext}
\begin{align}
    \Gamma_{\theta\theta} &= \langle\mathcal{O}(1)|\mathcal{O}(1)\rangle=\mathcal{O}(1)\\
    \Gamma_{\theta\alpha_2} &= \left.\left<\mathcal{O}(1)\right|A_2'(\alpha_2)\times\mathcal{O}(1) + B_2'(\alpha_2)\times\mathcal{O}(1)\right>=\epsilon_2'\times\mathcal{O}(1)\\
    \Gamma_{\theta\phi_2} &= \left.\left<\mathcal{O}(1)\right|A_2(\alpha_2)\times\mathcal{O}(1) + B_2(\alpha_2)\times\mathcal{O}(1)\right>=\epsilon_2\times\mathcal{O}(1)\\
    \Gamma_{\alpha_2\alpha_2} &= \left.\left<A_2'(\alpha_2)\times\mathcal{O}(1) + B_2'(\alpha_2)\times\mathcal{O}(1)\right|A_2'(\alpha_2)\times\mathcal{O}(1) + B_2'(\alpha_2)\times\mathcal{O}(1)\right>=(\epsilon_2')^2\times\mathcal{O}(1)\\
    \Gamma_{\alpha_2\phi_2} &= \left.\left<A_2'(\alpha_2)\times\mathcal{O}(1) + B_2'(\alpha_2)\times\mathcal{O}(1)\right|A_2(\alpha_2)\times\mathcal{O}(1) + B_2(\alpha_2)\times\mathcal{O}(1)\right>=\epsilon_2'\epsilon_2\times\mathcal{O}(1)\\
    \Gamma_{\phi_2\phi_2} &= \left.\left<A_2(\alpha_2)\times\mathcal{O}(1) + B_2(\alpha_2)\times\mathcal{O}(1)\right|A_2(\alpha_2)\times\mathcal{O}(1) + B_2(\alpha_2)\times\mathcal{O}(1)\right>=\epsilon_2^2\times\mathcal{O}(1),
\end{align}
\end{widetext}
as $\alpha_2 \to 0$, where
\begin{align}
    \epsilon_2'\equiv \begin{cases}
    A_2' & \text{if } B_2'=\mathcal{O}(A_2')\\
    B_2' & \text{if } A_2'=o(B_2')
    \end{cases},\label{eq:espilon2'}\\
    \epsilon_2\equiv \begin{cases}
    A_2 & \text{if } B_2=\mathcal{O}(A_2)\\
    B_2 & \text{if } A_2=o(B_2)
    \end{cases},
\end{align}
with $(A_2',B_2')$ denoting $\partial_{\alpha_2}(A_2,B_2)$, $x=\mathcal{O}(y)$ denoting $\mathrm{lim\,sup}_{\alpha_2\to0}(|x|/|y|)<\infty$, and $x=o(y)$ denoting $\mathrm{lim}_{\alpha_2\to0}(x/y)=0$. In short, $\epsilon_2'$ factors out of every row and column of $\Gamma$ that involves $\alpha_2$, while $\epsilon_2$ factors out of every row and column that involves a component of $\boldsymbol{\phi}_2$.

Then we observe that
\begin{align}\label{sigmasquared}
    \Sigma_{ij} = \left(\Gamma^{-1}\right)_{ij} = \frac{1}{|\Gamma|} C_{ij}
    \implies\rho_{ij}=\frac{C_{ij}}{\sqrt{C_{ii}C_{jj}}},
\end{align}
where $C$ is the cofactor matrix of $\Gamma$ (since $\Gamma$ is symmetric). From the above scaling relations, we have
\begin{align}
    C_{\theta\theta} &= (\epsilon_2')^2\epsilon_2^{2d}\times\mathcal{O}(1)\\
    C_{\theta\alpha_2} &= \epsilon_2'\epsilon_2^{2d}\times\mathcal{O}(1)\\
    C_{\theta\phi_2} &= (\epsilon_2')^2\epsilon_2^{2d-1}\times\mathcal{O}(1)\\
    C_{\alpha_2\alpha_2} &= \epsilon_2^{2d}\times\mathcal{O}(1)\\
    C_{\alpha_2\phi_2} &= \epsilon_2'\epsilon_2^{2d-1}\times\mathcal{O}(1)\\
    C_{\phi_2\phi_2} &= (\epsilon_2')^2\epsilon_2^{2d-2}\times\mathcal{O}(1),
\end{align}
where $d$ is the dimensionality of $\boldsymbol{\phi}_2$. It follows trivially from Eq.~\eqref{sigmasquared} that $\rho_{ij} = \mathcal{O}(1)$ for all $i,j$.

\bibliography{References}

\end{document}